# The Dual Impact of Artificial Intelligence in Healthcare: Balancing Advancements with Ethical and Operational Challenges

**Balaji Shesharao Ingole**
*IEEE member, Georgia, USA*

**Vishnu Ramineni**
*Albertsons Companies, Texas, USA*

**Nikhil Kumar Pulipeta**
*IEEE Senior member, Texas, USA*

**Manoj Jayntilal Kathiriya**
*IEEE Senior member, Connecticut, USA*

**Manjunatha Sughaturu Krishnappa**
*Oracle America inc, California, USA*

**Dr. Vivekananda Jayaram**
*IEEE Senior member, Texas, USA*

**Abstract**: *The synchronic and diachronic study of the evolution of Artificial Intelligence (AI) unveils one prominent fact that its effect can be traced in almost all fields such as healthcare industry. The growth is perceived holistically in software, hardware implementation, or application in these various fields. As the title suggests, the review will highlight the impact of AI on healthcare possibly in all dimensions including precision medicine, diagnostics, drug development, automation of the process, etc., explicating whether AI is a blessing or a curse or both. With the availability of enough data and analysis to examine the topic at hand, however, its application is still functioning in quite early stages in many fields, the present work will endeavour to provide an answer to the question. This paper takes a close look at how AI is transforming areas such as diagnostics, precision medicine, and drug discovery, while also addressing some of the key ethical challenges it brings. Issues like patient privacy, safety, and the fairness of AI decisions are explored to understand whether AI in healthcare is a positive force, a potential risk, or perhaps both.*

## INTRODUCTION

The world is transitioning in myriad ways and the role of AI in this entire process is undeniably great. Healthcare has been that sector where AI has brought tremendous revolution, from drug discovery, diagnosis and treatment, and clinical trials to other prospects. The medical field has explored several AI-driven applications that have somewhat eased out the tasks of healthcare workers, working as a transformative force reforming several aspects of the field.

AI has become a game-changer in healthcare. AI is offering new ways to diagnose, treat, and monitor patients. According to a study by Yu and colleagues [14], AI has already made significant improvements in patient care, healthcare, but the potential risks—like biased algorithms that can lead to unequal treatment that can't be ignored [6]. Obermeyer and Emanuel [13] have shown that if AI systems are trained on biased data, they could unintentionally deepen existing health disparities. Tackling these issues is key to ensuring that AI is used responsibly in healthcare. What is interesting to note here is that there is a discrepancy between healthcare expenditure, patient population, and the number of healthcare professionals. For instance, the overall healthcare expenditure in 2010 was around 6.9 billion (697 crores) which surged to 8 billion (800 crores) in 2024, while there seemed no sync in the number of medical professionals- for instance, there were 1.5 physicians per thousand patients which later barely increased to approximately 1.77 in 2024 [1] as shown in figure 1.

AI has the potential to reduce this disparity and offers a viable solution, especially by handling almost fifty percent of the tasks that were once managed by medical professionals such as personalized diagnostics, treatment planning, optimization of processes, surgery, and consultation. The present article aims to present both the risks and benefits of incorporating AI in the healthcare field. While AI can precisely identify tumours in medical images or diagnose rare diseases in no time thereby enhancing patient outcomes, it can also pose serious threats by involving data privacy and security risks, dangers of underdiagnosis, unequal treatment, and more. Here's an attempt to demonstrate a holistic picture incorporating both the positive and negative aspects to be considered before engaging in the discourse of assimilation of AI in healthcare, including the role of AI in revolutionizing the future of patient care [2], adhering to ethical standards and ensuring thoughtful implementation.

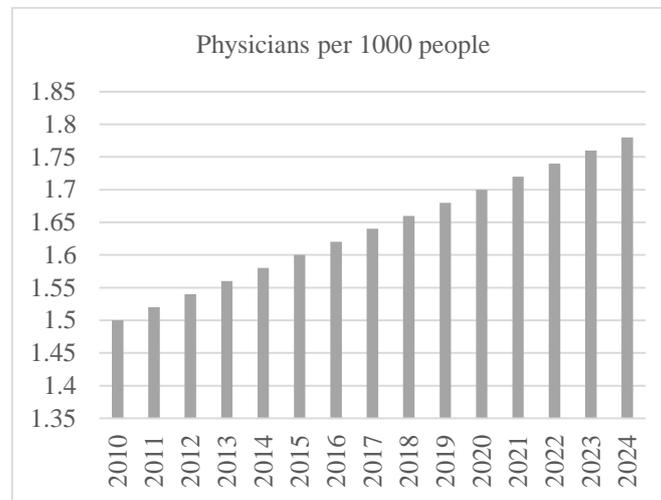

Figure 1.        Physicians Data

**Utilization of AI in Healthcare**

This portion of the paper will unfold multiple ways through which AI changes the entire dynamics of healthcare including enhanced diagnostic precision, increased drug discovery, and streamlined clinical trials. Its fundamental aids involve the identification of drug targets and compounds, profound interpretation of medical images for identifying diseases in no time, and management of large datasets for precisely analysing clinical trial outcomes [2]. In the first case i.e. identification of drug targets and compounds, AI accelerates the research and development process by examining biological data to identify the right molecule for effective treatments.

In the second case i.e. interpretation of medical images, AI enriches the accuracy rate of detecting a patient's medical condition that can't be easily grasped by the human eye [3]. In the third and final case i.e. management of large datasets for precisely analysing clinical trial outcomes, AI restructures the entire trial processes by optimizing the recruitment of patients, anticipating responses of the patients to medical treatments, and evaluating the trial outcomes.

From framing effective trial protocols to detecting issues at the early stage, AI-driven tools can assist in minimizing the duration of trials and enhance their success level [4]. However, AI can pose serious issues in the form of data breaches, over-dependence, and job losses (which will be explored extensively in the following discussion). With AI, the medical care of patients can be enriched exponentially as it helps in personalized treatment and remote monitoring quite effectively. Alongside the discussion of pros, it's essential to shed light on the challenges and the ways to fight them. For instance, one can consider employing robust and solid cybersecurity measures and maintaining a balance between healthcare professionals and AI technology integrated for a greater cause.

**Drug discovery**

The pace in the process of drug discovery or drug target identification is strengthened and hastened by the integration of AI technology in the healthcare field. AI can significantly process large biological datasets to find drug targets that incorporate genetic, protein, and metabolic data. For instance, Gupta R. reports that AI played an indispensable role in the identification of compounds for cancer treatment. Also, Pfizer's usage of IBM Watson, a computer system, was groundbreaking in pursuit of immune-oncology treatments [5] as shown in figure 2.

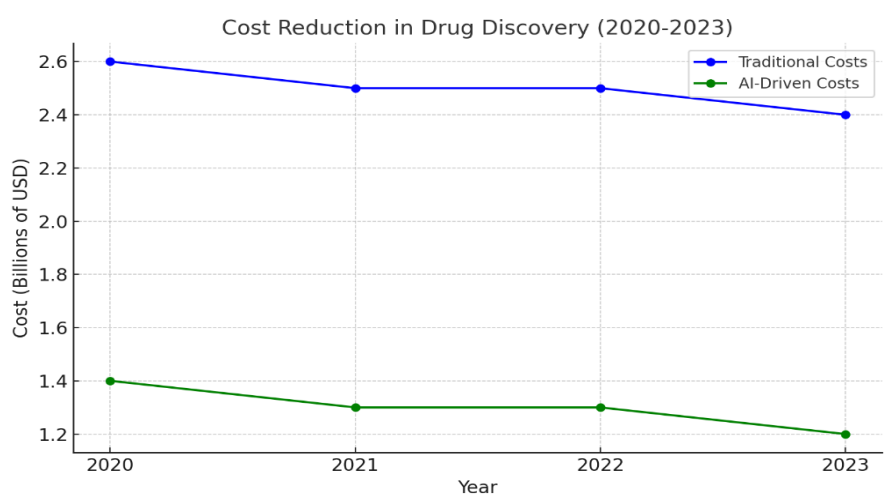

Figure 2.　　　Cost reduction in drug discovery

It must be noted that it's just the onset of a new era where, if the predictions are accurate, AI will bring revolution in the drug development process which will be relatively faster, affordable, efficient, and precise [6]. With the joint venture of pharmaceutical companies, tech industries, and academic institutions, AI can potentially improve drug discovery and assist scientists and researchers, but human intuition and creativity are irreplaceable phenomena.

There is significant cost reduction in drug discovery as shown in figure 3. There's a possibility of AI generating incorrect predictions if the data input is false or flawed and thus, overreliance on AI can prove to be detrimental. As AI gives correct output only when it's trained on correct data, therefore there are changes that AI may overlook the potential drug candidates if they don't fit into the patterns in the data, it has been trained on [7].

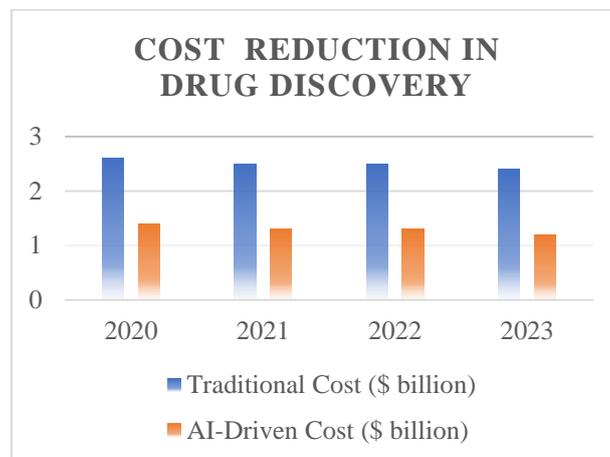

Figure 3.    Drug data

**Diagnosis and Treatment**

AI has proved its efficacy in the domain of healthcare by improving diagnostic precision and effectiveness [8]. There are subtle ways through which AI has shown its excellence, including interpreting medical images and distinguishing irregularities in radiological images accurately by using deep learning and computer vision. In the case of identifying life-threatening diseases such as pneumonia, cancer, eye diseases, etc., and reducing of time and cost involved in scans, integration AI in diagnostics and treatment can be beneficial as medical scans are stored and trained to ML algorithm.

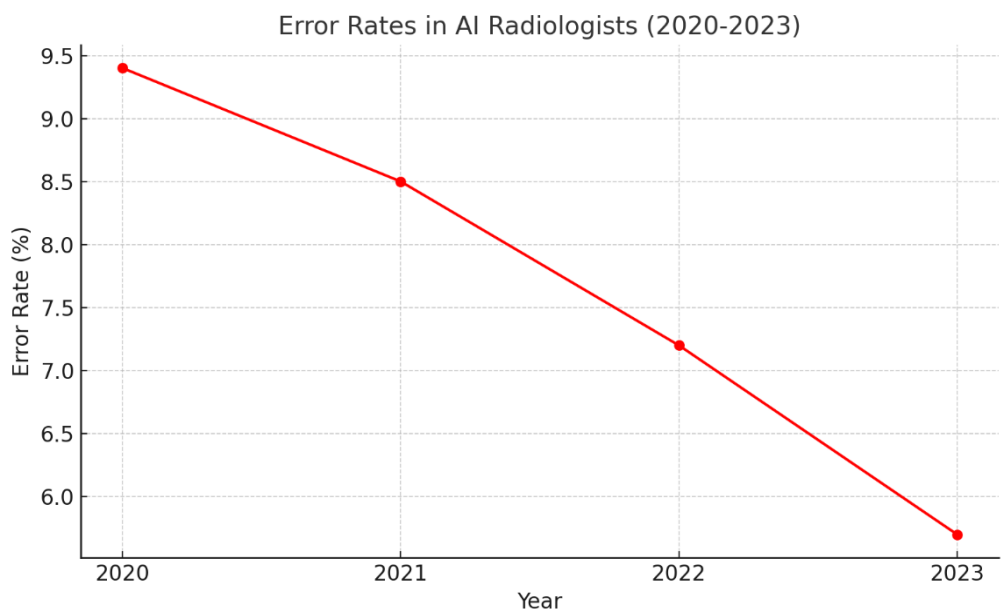

Figure 4.    Error Rates in AI Radiologists

Moreover, in the realm of pathology, AI's contribution is tremendous. There's a technique in medical image processing i.e. Pathology-Aware GAN-Bases Image

Augmentation that utilizes GAN (Generative Adversarial Networks) to produce realistic medical images showcasing health issues or diseases which will be later merged with original images to train computer systems to generate accurate medical images of patients. But what if any flawed input is given in the AI system, it might produce falsified images, risking patients' lives. One may encounter the danger of data breach as well. Healthcare professionals may highly rely on AI recommendations, eventually falling into the trap of falsification and error-making [9]. This results in reducing error rate as shown in figure 4 & 5.

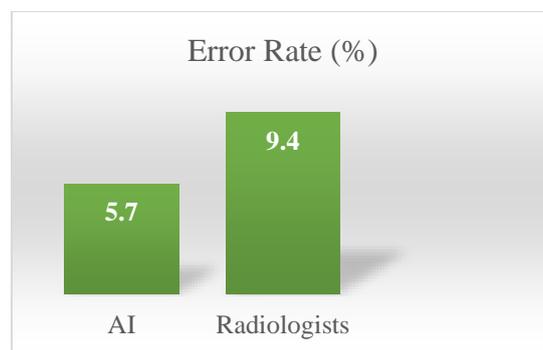

Figure 5.  Error rate data

**Clinical trial**

Among the two types of clinical research – clinical trials and observational trials – clinical trials entail the study of people's health and illness. Broadly speaking, these are research studies where multiple and distinct medical, surgical, and behavioural treatments are tested on people, unveiling the efficiency of new drugs or treatments on people [10]. Before the advent of AI, these clinical trials took a substantial amount of time, energy, and cost, surprisingly with lower success levels. AI changed the entire processes of healthcare, not only reducing the time through automated trials but also improving data monitoring facilities.

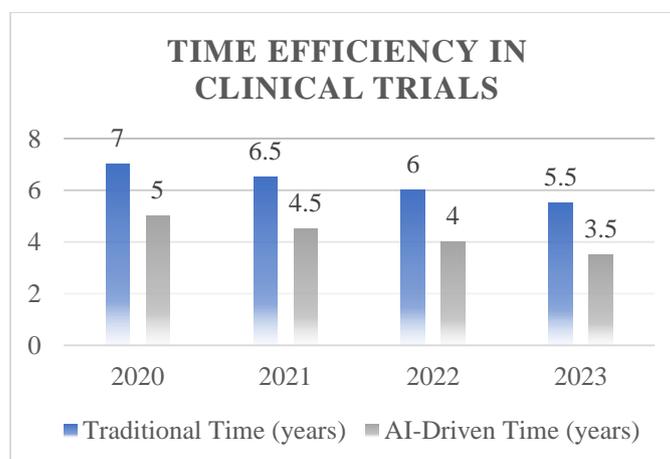

Figure 6.  Time efficiency data in clinical trials

However, as healthcare professionals and patients are both involved in the clinical trial, they both must show immense faith and trust in the systems. No one can predict its effect on patients' health and medical professionals' expertise and authority if AI offers challenges [15]. To address these challenges, the ethical codes and regulatory frameworks must be upgraded, guaranteeing that AI-driven clinical trials are conducted sensibly. Besides, educational and public awareness drives can instil trust and faith regarding the use of AI in clinical trials check stats in figure 6, enlightening the public about both the benefits and challenges.

**Clinical care**

Clinical care involves medical care and treatment of patients. AI identifies and diagnoses medical issues in patients and analyses their large data by utilizing medical imaging and machine learning. AI will be beneficial in managing necessary tasks such as scheduling appointments, predictive analytics, early intervention, and personalized preventive measures, maintaining medical records of patients, and recommending apt and personalized treatments for patients.

The cost of integration of AI in clinical care can be extremely higher than what you generally expect in areas such as infrastructure, training, monitoring, and more. Also, with the incessant growth of technology, healthcare regulations are bound to struggle to keep pace with it. Other issues such as patients' data privacy and security, healthcare professionals' job security, and health dilemmas among others are matters of serious concern [14]. These issues must be addressed through robust regulatory frameworks. To foster trust and ethics-based integration of AI in clinical care, there must be transparent communication about its risks and benefits and the potential efforts to eradicate dangers attached to its use. Moreover, with the proper engagement of patients, healthcare professionals, and policymakers, the incorporation of AI technology in clinical care can be made useful, ethical, and all-inclusive.

**AI IN PRECISION MEDICINE**

The evolution of AI from rule-based systems to robust Machine Learning and Deep Learning algorithms has been profound, especially in the field of healthcare. The amalgamation of AI and precision medicine is beneficial as it helps in the revolutionization of patient care by offering personalized treatment plans, enriching diagnostic aptness, revamp disease cure procedures. Several techniques, for medical imaging and generating synthetic medical images, such as Convolutional Neural Networks (CNNs) and Generative Adversarial Networks (GANs) are utilized. For the analysis of patient data and anticipation of disease progression, Boltzmann Machines, K Nearest Neighbour (KNN), and Recurrent Neural Networks (RNNs) are utilized.

As mentioned already one of the main purposes of the paper is to trace the evolution of AI (in the healthcare sector) from generalized to personalized medicine where treatment will be tailored based on the specific requirement of every patient and their data including genetics, medical history, and way of living. The operational efficacy can also be enhanced by AI as it analyses large datasets to comprehend specific patterns that might be missed by the human eye. AI can complement healthcare professionals in making sound decisions and improve the treatment process, for instance, AI in radiotherapy planning can optimize the quality and efficacy of the treatment process. It is highly emphasized that legislation and ethical frameworks must be kept as a priority to ascertain AI's universality and reliance in clinical practice [11].

**AI ROBOTICS ASSISTED SURGERIES**

The entire phenomenon of surgery is transformed by the introduction of AI Robot-assisted Surgery (RAS) that provides exposure to healthcare professionals to robust applications made to increase surgical accuracy, anticipate results, and augment surgical training. AI application that monitors motion analysis assists in understanding the intricacies involved in surgeries and heighten the precision of robotic tools. Researches demonstrate that AI has proven its usefulness not only in identifying surgical instruments and predicting results such as urinary continence and recovery time after the Robot-assisted Radical Prostatectomy (RARP) process but also in handling complex and intricate procedures involved in surgical fields like urology and gynaecology [13].

Nevertheless, it's important to highlight the limitations related to relatively small sample sizes and scarcity of standardization in AI algorithms and datasets which later leads to compromise in patients' safety and potential surgical outcomes. The social, legal, and ethical concerns must also be considered diligently. Amidst such challenges, AI can outshine if future researchers focus more on conducting extensive studies, validating AI algorithms externally, and guaranteeing transparency for both medical surgeons and patients so that there remains no communication gap [12].

**ECONOMIC IMPACT OF AI ON HEALTHCARE**

It's cited above that the implementation of AI in the healthcare sector boosts its economic stability by ensuring cost reduction at an unprecedented rate. In the U.S. the annual estimated saving percentage is between 5-10% which is equivalent to $200-$360 billion [16]. With optimization of operating room schedules, streamlining of supply chains, and management of administrative works, AI's assistance in minimizing overall healthcare expenses. Researchers claim that AI technologies such as machine learning (ML) or natural language processing etc. are promoting patient outcomes and prolific satisfaction (of both healthcare professionals and patients) by ensuring medical imaging accuracy, and precision in predictive analytics for clinical care of patients, and automation of documents processing in electronic medical archives.

Democratization of healthcare facilities can be done with AI offering robust diagnostic tools and individualized treatment procedures, protecting, or serving those who are underserved. We must be reminded that prevention is better than cure, therefore, early detection and treatment seem more fruitful than interventions in the later stage when things might get unmanageable. AI can potentially overhaul the insurance domain by managing claims and mitigating errors, leading to savings of $80- $110 billion per year. Hence, the claim process becomes easy and effective for both the parties- medical professionals and patients- making healthcare more accessible and inexpensive [17].

Major challenges such as reliance on traditional payment models, paucity of structured datasets, and trained personnel are slowing down the process of AI's integration in healthcare. A proper strategy that incorporates the preparation of an aim-based roadmap, adoption of agile delivery models, implementation of apt technologies, and restructuring of healthcare operations, will be required to overcome these challenges smoothly.

**WHAT'S THE FUTURE OF AI IN HEALTHCARE?**

AI is an ever-evolving phenomenon and thus, its impact on several domains is profoundly engaging from both positive and negative sides. AI is known for automation, fastened processes, and less human input which could potentially revolutionize several industries including healthcare and city planning [18]. For instance, in the healthcare field, AI can offer personalized treatment procedures, systematic analysis of patient health datasets, accurate diagnostic analytics, and prediction of illnesses; in smart city planning, AI can heighten infrastructure efficacy, manage traffic chaos, minimize energy consumption, and ensure public safety.

Besides these pros, AI's integration can raise serious concerns from ethical and societal perspectives. The autonomy of AI may jeopardize the basic codes and norms of society, and thus, it becomes necessary to keep AI's decision-making processes in sync with societal values and conduct. Overcoming challenges such as biases that result in unjust and unfair practices, a data breach that endangers patients' privacy, job losses, etc. become inevitable concern [19].

**TABLE I: Benefits and Risks of AI Applications in Healthcare**

| Sr. No. | AI Application | Benefits | Potential Risks |
|---|---|---|---|
| 1. | Drug Discovery | Faster, more cost-effective drug development | Over-dependence on data, flawed prediction. |
| 2. | Medical Imaging | Improved accuracy in diagnosing diseases | Incorrect diagnosis |
| 3. | Clinical trials | Accurate results and less time and cost | Trust issues, health risks |
| 4. | Clinical care | Efficient management | Job loss |

The future of AI will require a balanced, informed, and strategic approach to eradicate these challenges, ensuring transparency, inclusivity, and ethics-based development. There must be congenial collaboration among national and international stakeholders to set up global standards and frameworks that can ensure responsible and ethical deployment of AI technologies in healthcare. Such deployment must be always under surveillance so that no one's life is at stake.

Table I demonstrates the extraordinary benefits and potential challenges of AI applications in healthcare facilities such as drug discovery, medical imaging, clinical trials, and clinical care.

**CONCLUSION**

The inclusion of AI in healthcare will not only be a technological boost but also a thoughtful take on patient care and welfare. By leveraging AI, there'll be an exponential surge in diagnostic precision, treatment results, and personalized attention to each patient, making it more patient-centric. The amalgamation of AI and precision medicine is beneficial as it helps in the revolutionization of patient care by offering personalized treatment plans, enriching diagnostic aptness, revamp disease cure procedures. As AI continues to grow in healthcare, it brings both incredible opportunities and complex challenges. While AI can improve patient care, reduce costs, and make diagnoses more accurate, we must address critical ethical concerns like data privacy, algorithmic bias, and the need for transparency. The future of AI in healthcare will depend on how well we manage these challenges, ensuring that AI remains a tool to enhance human care rather than replace it. Collaboration among healthcare professionals, AI developers, and policymakers will be key in making sure AI is used responsibly and effectively [6].

**REFERENCES**

[1] A. Blanco-González et al., "The Role of AI in Drug Discovery: Challenges, Opportunities, and Strategies," Multidisciplinary Digital Publishing Institute (MDPI), vol. 16, no. 6, pp. 891-900, Jun. 2023. doi: 10.3390/ph16060891.
[2] A. A. Abujaber and A. J. Nashwan, "Ethical framework for artificial intelligence in healthcare research: A path to integrity," World J Methodol, vol. 14, no. 3, pp. 940-971, Sep. 2024. doi: 10.5662/wjm.v14.i3.94071.
[3] M. Yousef Shaheen, "Applications of Artificial Intelligence (AI) in healthcare: A review," vol. 1, pp. 1-8, 2021. doi: 10.14293/S2199-1006.1.SOR-.PPVRY8K.v1.
[4] H. E. Kim et al., "Changes in cancer detection and false-positive recall in mammography using artificial intelligence: A retrospective, multireader study," Lancet Digit Health, vol. 2, no. 3, pp. e138–e148, Mar. 2020. doi: 10.1016/S2589-7500(20)30003-0.
[5] K. K. Mak and M. R. Pichika, "Artificial intelligence in drug development: Present status and future prospects," Drug Discovery Today, vol. 24, no. 3, pp. 773-780, Mar. 2019. doi: 10.1016/j.drudis.2018.11.014.
[6] A. M. Challen et al., "Artificial intelligence, bias and clinical safety," BMJ Quality & Safety, vol. 28, no. 3, pp. 231-237, Mar. 2019. doi: 10.1136/bmjqs-2018-008370.

[7] B. Yadav Kasula, "Advancements in AI-driven Healthcare: A Comprehensive Review of Diagnostics, Treatment, and Patient Care Integration," ResearchGate, 2023. Available: https://www.researchgate.net/publication/378314433.

[8] D. Ueda et al., "Fairness of artificial intelligence in healthcare: Review and recommendations," Springer, vol. 14, no. 4, pp. 74-91, Jan. 2024. doi: 10.1007/s11604-023-01474-3.

[9] C. H. Wong, K. W. Siah, and A. W. Lo, "Estimation of clinical trial success rates and related parameters," Biostatistics, vol. 20, no. 2, pp. 273–286, Apr. 2019. doi: 10.1093/biostatistics/kxx069.

[10] S. Kumar Chintala, "AI-driven personalised treatment plans: The future of precision medicine," Journal of Precision Medicine, vol. 17, no. 2, pp. 33-45, 2023.

[11] A. Moglia, K. Georgiou, E. Georgiou, R. M. Satava, and A. Cuschieri, "A systematic review on artificial intelligence in robot-assisted surgery," International Journal of Surgery, vol. 89, pp. 1-10, Nov. 2021. doi: 10.1016/j.ijsu.2021.106151.

[12] E. J. Topol, *Deep Medicine: How Artificial Intelligence Can Make Healthcare Human Again*, New York: Basic Books, 2019.

[13] Z. Obermeyer and E. J. Emanuel, "Predicting the future—Big data, machine learning, and clinical medicine," New England Journal of Medicine, vol. 375, no. 13, pp. 1216-1219, Sep. 2016. doi: 10.1056/NEJMp1606181.

[14] K.-H. Yu, A. L. Beam, and I. S. Kohane, "Artificial intelligence in healthcare," Nature Biomedical Engineering, vol. 2, no. 10, pp. 719-731, Oct. 2018. doi: 10.1038/s41551-018-0305-z.

[15] D. A. Hashimoto, G. Rosman, D. Rus, and O. R. Meireles, "Artificial intelligence in surgery: Promises and perils," Annals of Surgery, vol. 268, no. 1, pp. 70-76, Jul. 2018. doi: 10.1097/SLA.0000000000002693.

[16] A. Esteva et al., "Dermatologist-level classification of skin cancer with deep neural networks," Nature, vol. 542, no. 7639, pp. 115-118, Feb. 2017. doi: 10.1038/nature21056.

[17] F. Jiang et al., "Artificial intelligence in healthcare: Past, present, and future," Stroke and Vascular Neurology, vol. 2, no. 4, pp. 230-243, Dec. 2017. doi: 10.1136/svn-2017-000101.

[18] A. N. Ramesh, C. Kambhampati, J. R. T. Monson, and P. J. Drew, "Artificial intelligence in medicine," Annals of the Royal College of Surgeons of England, vol. 86, no. 5, pp. 334-338, Sep. 2004. doi: 10.1308/147870804290.

[19] E. J. Topol, "A decade of digital medicine innovation," Nature Medicine, vol. 26, no. 1, pp. 8-9, Jan. 2020. doi: 10.1038/s41591-019-0729-5.